\documentclass[journal,onecolumn,12pt,twoside]{IEEEtran}

\normalsize

\usepackage{cite}
\usepackage[cmex10]{amsmath}
\usepackage{amsfonts,amssymb,amscd,amsthm,xspace,amstext,makeidx}
\usepackage{algorithmic}
\usepackage{array}
\usepackage{mdwmath,mdwtab,eqparbox}
\usepackage{graphicx}
\usepackage[tight,footnotesize]{subfigure}

\normalsize
\theoremstyle{plain}

\newtheorem{theorem}{Theorem}
\newtheorem{corollary}[theorem]{Corollary}

\newtheorem{proposition}[theorem]{Proposition}

\theoremstyle{remark}

\newcommand{\fref}[1]{Fig.~\ref{#1}}

\newcommand{\eref}[1]{(\ref{#1})}

\newcommand{\cref}[1]{Chapter~\ref{#1}}
\newcommand{\sref}[1]{\S-\ref{#1}}

\newcommand{\di}{\mathrm{d}}

\newcommand{\erfc}{\mathrm{erfc}}
\newcommand{\one}{\mathbf{1}}

\newcommand{\E}{\mathbf{E}}
\newcommand{\exu}{\mathcal{E}_{\mathrm{U}}}
\newcommand{\exd}{\mathcal{E}_{\mathrm{D}}}
\newcommand{\Pb}{\mathbf{P}}

\newcommand{\dc}{d_{\textrm{c}}}
\newcommand{\tauu}{\tau_{\textrm{u}}}
\newcommand{\taud}{\tau_{\textrm{d}}}

\begin{document}

\title{Some Properties of Large Excursions of a Stationary Gaussian Process}
\author{Van~Minh~Nguyen
\thanks{V. M. Nguyen is with the R\&D Department of Sequans Communications, 19 Parvis de la D\'{e}fense, 92073 Paris-La D\'{e}fense Cedex, France.
email: vanminh.nguyen@sequans.com.}}


\maketitle

\begin{abstract}
\boldmath
    The present work investigates two properties of level crossings of a stationary Gaussian process $X(t)$ with autocorrelation function $R_X(\tau)$. We show firstly that if $R_X(\tau)$ admits finite second and fourth derivatives at the origin, the length of up-excursions above a large negative level $-\gamma$ is asymptotically exponential as $-\gamma \to -\infty$. Secondly, assuming that $R_X(\tau)$ admits a finite second derivative at the origin and some defined properties, we derive the mean number of crossings as well as the length of successive excursions above two subsequent large levels. The asymptotic results are shown to be effective even for moderate values of crossing level. An application of the developed results is proposed to derive the probability of successive excursions above adjacent levels during a time window.
\end{abstract}



\section{Introduction}

    Crossings of a stochastic process $X(t)$ with respect to a certain level represent the nature of a wide range of theoretical problems as well as practical applications. It is a subject that has received much study for long time, and has been known as \emph{level crossing theory}. Its first milestone dates back to the middle of the last century with the pioneering works of Rice \cite{Rice1945,Rice1958}, which have motivated numerous fruitful investigations on this subject. The fundamental results were quite fully summarized in \cite{Cramer1967,Leadbetter1983}.

    This classical topic continues to attract attention in the research community, especially from applied sciences, and has regularly received recent contributions, see e.g. \cite{Kratz2006} for a survey. In particular, some recent investigations in the communication and networking domain include \cite{Blachman1996,Blachman1998} where Blachman investigated the shape of excursions and peak values of a frequency-modulated signal, \cite{Morgan2007} where Morgan addressed the level crossings of a discrete-time process and showed its connection to Rice's fundamental results for a continuous-time process, and \cite{Ramos-Alarcon2009} where Ramos-Alarcon et al. studied the level crossing duration distribution of a Nakagami fading process. Beside these theoretically oriented works, applications in \cite{Vijayan1993a,Mandayam1998,Graziosi1999,Jiang2003a,Yang2004,Nguyen2011Icc} have confirmed clear interest of this theory in wireless communications.

    The source of motivation behind the present paper also originated from problems related to the crossings of radio signals with respect to some threshold. To define our position within this rich literature, we first present a short summary of related results. In  \cite{Rice1945,Rice1958}, Rice firstly provided the closed-form expression of the average crossing rate, and proved that under favorable conditions, excursions of a sample path above a large level behave asymptotically as a parabola. The latter has been referred to as \emph{large excursions} in the literature. The technical conditions of the theorem have been weakened by, among others, Ivanov \cite{Ivanov1960} and Ylvisaker \cite{Ylvisaker1965}. Also related to this property, Blachman \cite{Blachman1988} refined the parabolic approximation for large intervals. On the other hand, the time interval between two consecutive up-crossings of a level was proved to asymptotically follow an exponential distribution when the level is large, see \cite[\S 12.4]{Cramer1967}. Also from \cite[p. 258]{Cramer1967}, it was shown that the time instants of up-crossings of successive levels form a Poisson point process.

    Originally motivated by the analysis of radio link failure and measurement triggering in mobile cellular systems (which can be modeled as excursions, and as successive excursions of the signal with respect to some thresholds), this paper is aimed at exploiting two of the open questions of this theory. The first is related to \emph{how up-excursions behave above a large negative level}, referred to as \emph{large negative excursions}. Our answer to this question forms a complement to the asymptotic parabolic property of large excursions established by Rice \cite{Rice1945,Rice1958}. The second question concerns \emph{asymptotic properties of successive crossings and successive excursions of two subsequent large levels}. The results pertaining to this question provide another view and a complement to the Poisson point process property of successive crossings as cited above.

    We investigate the above two questions by considering a stationary normal process $X(t)$. Under appropriate conditions specified for the autocorrelation function $R_X(\tau)$ of $X(t)$, and based on the exponential distribution of the interval between two consecutive up-crossings, we show in \sref{Sec:ExcurSmallLevel} that the length of low-level excursions above level $-\gamma$ is asymptotically an exponential distribution with rate equal to the up-crossing rate as $-\gamma \to -\infty$. After that, using the parabolic trajectory property of large excursions, we develop in \sref{Sec:ExcurMultLevel} some properties associated with crossings of $X(t)$ of two successive large levels $\gamma_1$ and $\gamma_2 \geq \gamma_1$. We obtain the mean number of crossings of $X(t)$ of level $\gamma_2$, and the distribution of the length of up-excursions above $\gamma_2$, given that $X(t)$ has an up-excursion above $\gamma_1$ as $\gamma_1 \to +\infty$. An application using the probability of successive excursions of $X(t)$ during some given time window is then derived in \sref{Ssec:ExcurMultLevelApp} to show the usefulness of the developed results.

\section{Notation}

    Let $X(t)$ be a real-valued stationary Gaussian process of continuous parameter $t$, and have \emph{zero mean} and \emph{unit variance}. Write $R_X(\tau)$ as the autocorrelation function of $X(t)$. In addition, we assume that $X(t)$ is ergodic so that the properties of $X(t)$ can be studied by those of its sample path. For notational simplicity, let us use $X(t)$ to refer to a sample path. Assume that $X(t)$ is not identically equal to any fixed level $\gamma$ during any non-empty interval of $t$. Let us describe some basic notations which can be found in \cite{Cramer1967}.

    We say that $X(t)$ has an up-crossing of the level $\gamma$ at $t_0$ if there exists $\epsilon > 0$ such that $X(t) \leq \gamma$ for $t \in (t_0 - \epsilon, t_0)$, and $X(t) \geq \gamma$ for $t \in (t_0, t_0+\epsilon)$. A down-crossing of $X(t)$ is similarly defined by reserving inequalities in the above definition. Intuitively, since $X(t)$ is assumed continuous and not identically equal to $\gamma$ in any subinterval, an up-crossing, a down-crossing, respectively, of the level $\gamma$ at $t_0$ is described by the fact that $X(t)-\gamma$ changes sign from non-positive to non-negative, from non-negative to non-positive, respectively, when $t$ goes from a left to a right neighborhood of $t_0$. And we say that $X(t)$ has a crossing of the level $\gamma$ at $t_0$ if $X(t_0) = \gamma$ and there exist $t_1$ and $t_2$ in a neighborhood of $t_0$ 
    such that $[X(t_1) - \gamma][X(t_2) - \gamma] < 0$.

    Using the above definition of crossings, we say that $X(t)$ has an up-excursion above level $\gamma$ during $[t_1, t_2]$ if $X(t)$ has an up-crossing of $\gamma$ at $t_1$, then a down-crossing of $\gamma$ at $t_2$, and does not have any crossing of $\gamma$ during $(t_1, t_2)$. A down-excursion below level $\gamma$ is similarly defined.

    Before describing some fundamental results, let us note that the condition $X(t) \geq \gamma$ is equivalent to $-X(t) \leq -\gamma$. As $X(t)$ is Gaussian and centered at zero, $-X(t)$ is also a Gaussian process statistically identical to $X(t)$. Therefore, the properties of up-crossings, up-excursions of $X(t)$ with respect to level $\gamma$ are directly applicable to down-crossings, down-excursions of $X(t)$ with respect to level $-\gamma$.

\section{Related Results}

\begin{figure}
        \centering
        \subfigure[]
        {
            \includegraphics[width=0.45\textwidth]{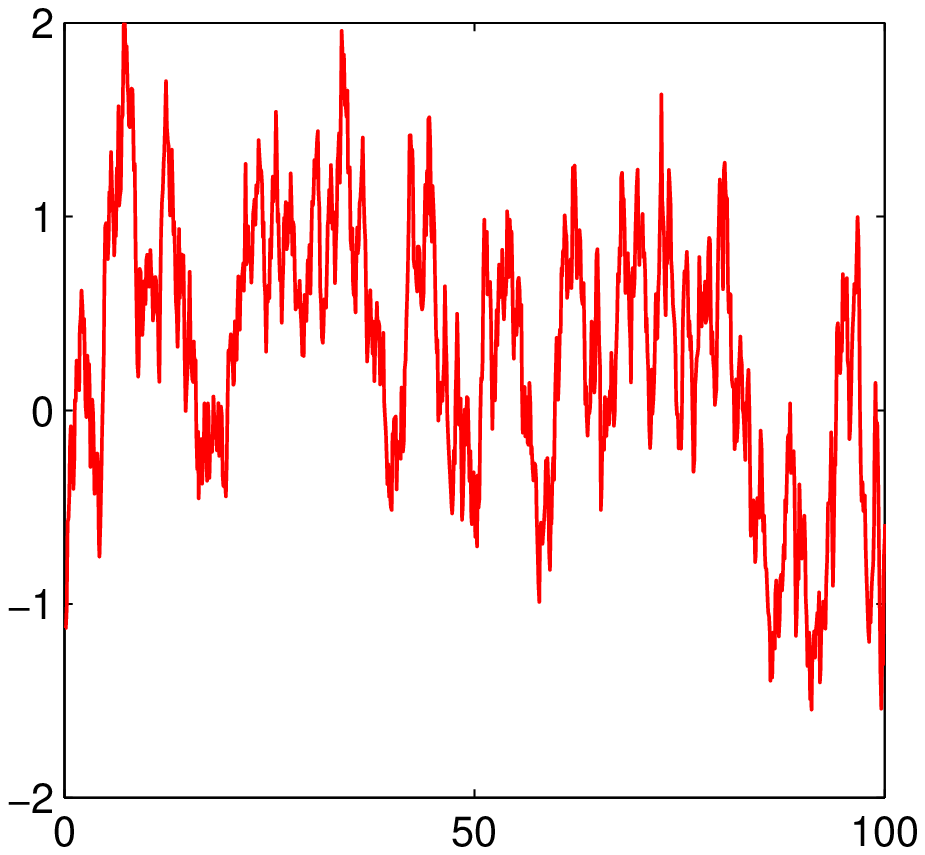}
        }
        \subfigure[]
        {
            \includegraphics[width=0.45\textwidth]{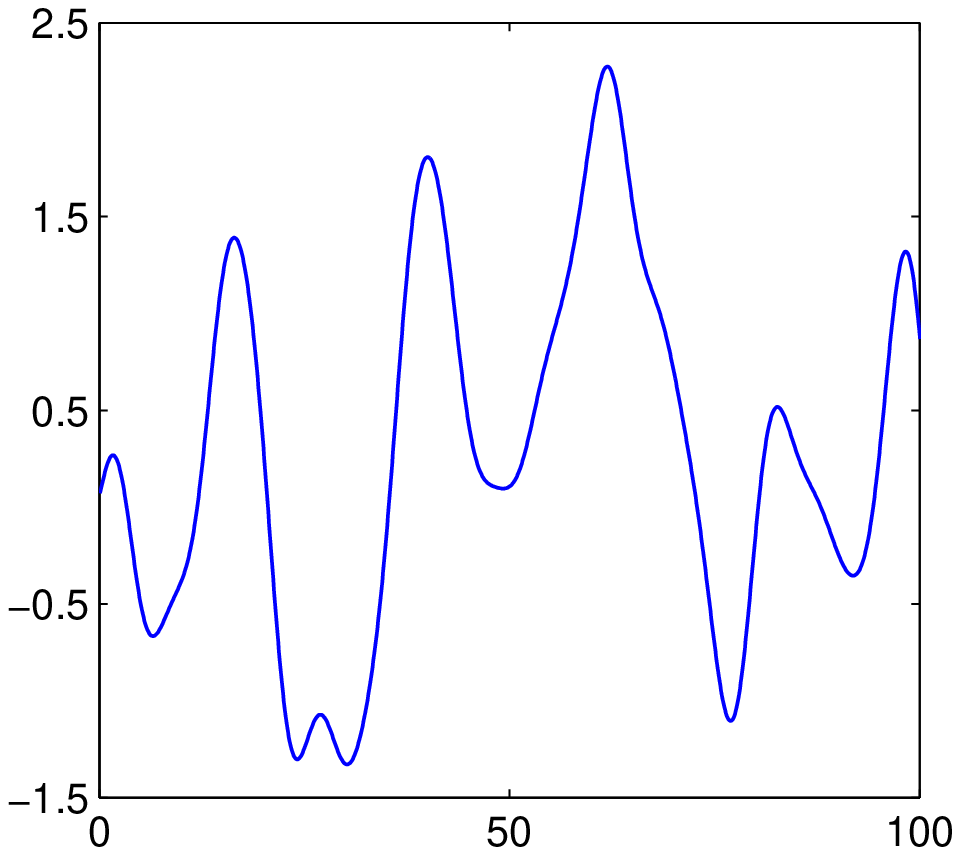}
        }
        \caption{Sample paths of $X(t)$ simulated with $\dc=5$ and discrete step size 0.1. (a) Gauss-Markov process. (b) Gaussian process with squared exponential autocorrelation function.}\label{Fig:X}
    \end{figure}

    Write $C_{\gamma}$ the number of crossings of $X(t)$ of the level $\gamma$ during a unit time interval.
    \begin{theorem}[Cram\'er and Leadbetter \cite{Cramer1967}]\label{Thm:LCR}
        With the notation developed and $X(t)$ defined as above:
        \begin{equation}\label{Eq:Bgd:LCR}
            \E C_{\gamma} = \frac{1}{\pi}\sqrt{\frac{\lambda_2}{\lambda_0}}\exp\left(-\frac{\gamma^2}{2\lambda_0}\right),
        \end{equation}
        where $\lambda_0 = R_X(0)$ and $\lambda_2 = - R_X''(\tau)|_{\tau = 0}$, and where $\E C_{\gamma} < +\infty$ if and only if $\lambda_2 < +\infty$.
    \end{theorem}
    The above last condition is equivalently stated that the level crossing rate is finite if and only if the autocorrelation function $R_X(\tau)$ of $X(t)$ has finite second derivative at the origin. According to Leadbetter et al. \cite{Leadbetter1983}, this condition is satisfied if $R_X$ admits the following form:
    \begin{equation}\label{Eq:ACF_Cond1}
        R_X(\tau) = 1 - \frac{\lambda_2 \tau^2}{2} + o(\tau^2), \quad \textrm{ as } \tau \to 0,
    \end{equation}
    with finite $\lambda_2$. To illustrate this property, \fref{Fig:X} plots sample paths of two processes:
    \begin{itemize}
    \item[-] \emph{Gauss-Markov process with exponential autocorrelation function \footnote{This process can be expressed as $X(n+1) = a X(n) + \mathcal{N}(0, \sigma_n)$, which refers to the Markov property.}:}
        \begin{equation} \label{eq:acfmarkov}
            R_X(\tau) = \exp\left(-\frac{|\tau|}{\dc}\right),
        \end{equation}
    \item[-] \emph{Gaussian process with squared exponential autocorrelation function:}
        \begin{equation}\label{eq:acfdblexp}
            R_X(\tau) = \exp\left(-\frac{1}{2}(\frac{\tau}{\dc})^2\right),
        \end{equation}
    \end{itemize}
    where $\dc$ is a positive constant that can be roughly thought of as the correlation distance in time that we have to move in order to observe significant change of $X(t)$. The autocorrelation function $R_X(\tau)$ of the Gauss-Markov model is not differentiable, while the squared exponential autocorrelation function is infinitely differentiable. We can see in \fref{Fig:X} that \eref{eq:acfdblexp} results in a very smooth process, while \eref{eq:acfmarkov} results in a lot of fluctuations. These rapid fluctuations illustrate the infinity of the mean crossing rate.

    Let $U_{\gamma}$ and $D_{\gamma}$ be the number of up-crossings and down-crossings of $X(t)$ of the level $\gamma$ during a unit time period.
    \begin{proposition}[Cram\'er and Leadbetter \cite{Cramer1967}]\label{Prop:UpDownCrossRate}
        With the same assumptions for the process $X(t)$, if $\lambda_2 < +\infty$:
        \begin{equation}
            \E U_{\gamma} = \E D_{\gamma} = \E C_{\gamma}/2.
        \end{equation}
    \end{proposition}
    This means that when $\lambda_2 < +\infty$, the level crossing rate is equally shared between the up-level and the down-level crossing rates.

\begin{figure}
        \centering
        \subfigure[PDF]
        {
            \includegraphics[width=0.45\textwidth]{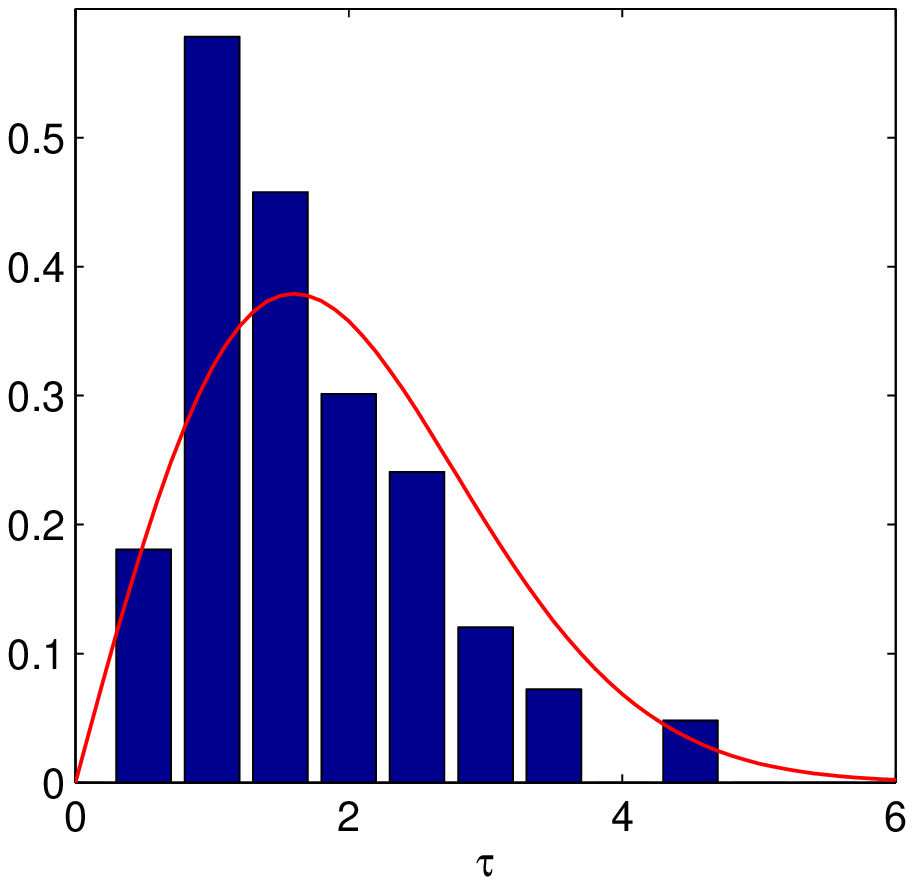}
        }
        \subfigure[CDF]
        {
            \includegraphics[width=0.45\textwidth]{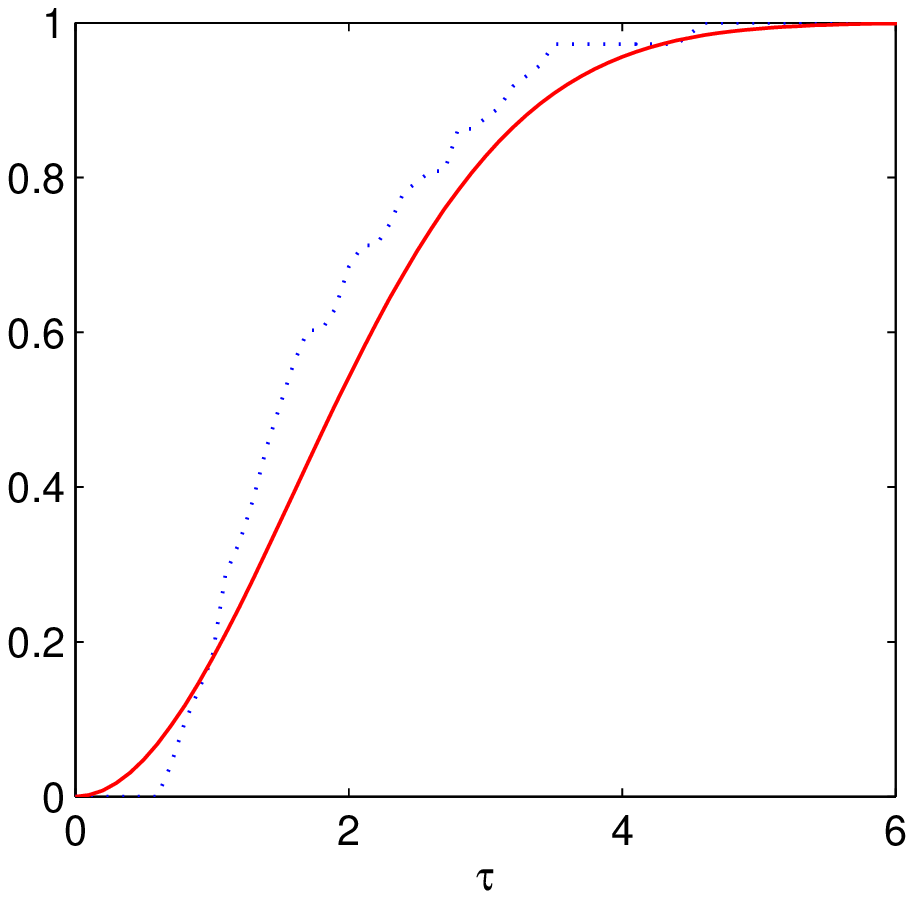}
        }
        \caption{Length of excursions above a large level $\gamma$ evaluated with squared exponential ACF with $\dc = 2$, and $\gamma = 2.5$. (a) Plot of Rayleigh pdf (solid) against the histogram. (b) Plot of Rayleigh cdf (solid) against the normalized cumulated sum of the histogram.}\label{Fig:larExcur}
    \end{figure}

    We are now interested in another result related to crossings of a large level $\gamma \to \infty$. In the following, we provide results related to an up-crossing of a large level; the corresponding results of down-crossings of a large negative level will be directly obtainable by the aforementioned symmetry property.
    \begin{theorem}[Thm. 10.4.2 \cite{Leadbetter1983}]\label{Thm:AsympExcursion}
        With the process $X(t)$ described above, if its autocorrelation function $R_X(\tau)$ satisfies
        \begin{equation}\label{Eq:ACF_Cond2}
            R_X''(\tau) = \lambda_2 + O(|\log|\tau||^{-a}) \quad \textrm{ as } \tau \to 0
        \end{equation}
        with finite $\lambda_2$ for some $a > 1$, and
        \begin{equation}\label{Eq:ACF_Cond3}
        R_X(\tau) \to 0 \quad \textrm{ as } \tau \to +\infty,
        \end{equation}
        then, as $\gamma \to +\infty$, excursions of $X(t)$ above $\gamma$ behave asymptotically as
        \begin{equation}\label{Eq:AsympParabola}
            X(t) \sim \gamma + \xi t - \gamma \frac{\lambda_2 t^2}{2},
        \end{equation}
        where $\xi$ is a Rayleigh random variable of parameter $\sqrt{\lambda_2}$.
    \end{theorem}
    Intuitively, trajectories of $X(t)$ above a large level $\gamma$ behave asymptotically as parabolas with Rayleigh distributed parameter $\xi$. \fref{Fig:larExcur} shows that the asymptotic distribution of the length of large excursions matches with simulation results.

    The next related result deals with the interval between an up-crossing and the $k$th subsequent up-crossing for $k = 1,2,\ldots$. Consider the above stationary normal process $X(t)$ such that its autocorrelation function $R_X(\tau)$ satisfies the following two additional assumptions (\cite[\S 12.1]{Cramer1967}):
    \begin{equation}\label{Eq:ACF_Cond4}
        R_X(\tau) = 1 - \frac{\lambda_2}{2!}\tau^2 + \frac{\lambda_4}{4!}\tau^4 + o(\tau^4)
    \end{equation}
    with finite $\lambda_2$ and $\lambda_4$, as $\tau \to 0$, and
    \begin{equation}\label{Eq:ACF_Cond5}
        R_X(\tau) = O(\tau^{-a})
    \end{equation}
    for some $a > 0$, as $\tau \to +\infty$. The condition \eqref{Eq:ACF_Cond4} implies that $X(t)$ has, with probability 1, a continuous sample function derivative, and the condition \eqref{Eq:ACF_Cond5} implies that the spectrum of $X(t)$ is everywhere continuous so that $X(t)$ is ergodic, see \cite[\S 12.1]{Cramer1967}. Let $F_{k}(t)$ be the distribution of the interval $t$ between an up-crossing and the $k$th subsequent up-crossing, and denote $\mu = \E U_{\gamma}$.
    \begin{theorem}[\cite{Cramer1967}]\label{Thm:IntervalBetweenUpcrosses}
        With the above assumptions and notation:
        \begin{equation}
        \lim_{\gamma \to \infty}F_{k}\left(\frac{t}{\mu}\right) = 1 - \left[\sum_{n = 1}^k \frac{t^{n-1}}{(n-1)!}\right]e^{-t}, \quad k = 1, 2, \ldots.
        \end{equation}
        The probability density function of this limiting distribution is
        \begin{equation}
            f_k\left(\frac{t}{\mu}\right) = \frac{t^{k-1}}{(k-1)!} e^{-t}
        \end{equation}
        with mean $k$.
    \end{theorem}

\begin{figure}
        \centering
        \subfigure[PDF]
        {
            \includegraphics[width=0.45\textwidth]{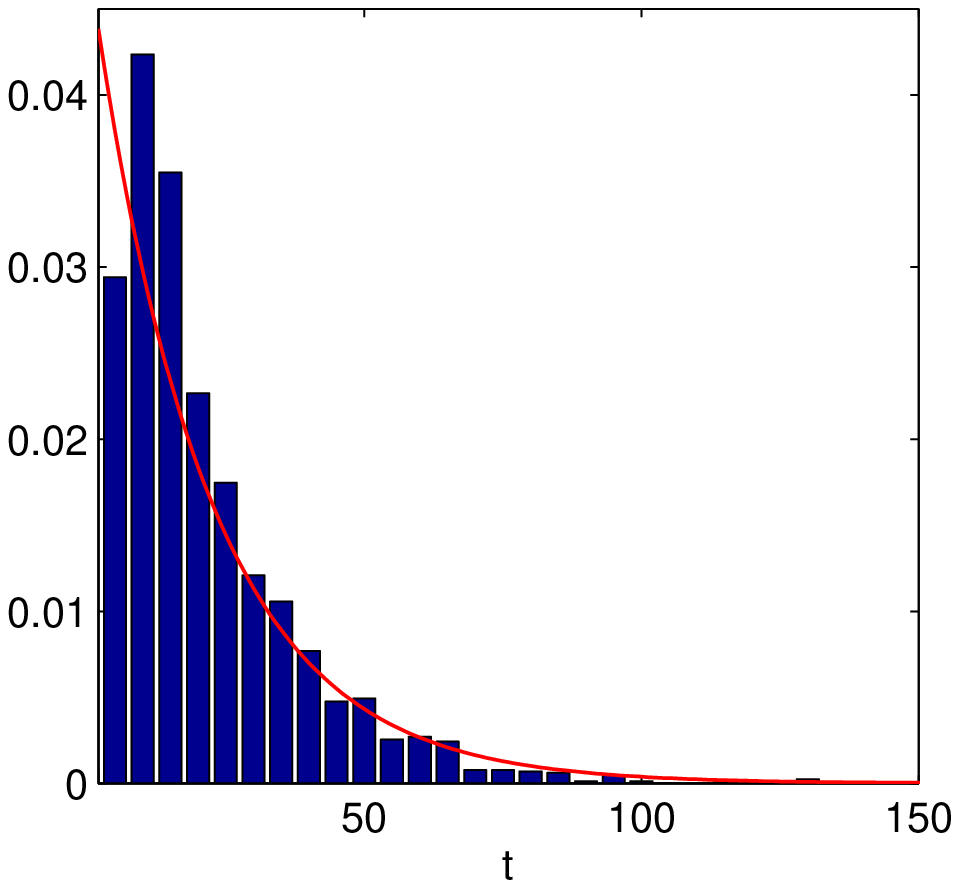}
        }
        \subfigure[CDF]
        {
            \includegraphics[width=0.45\textwidth]{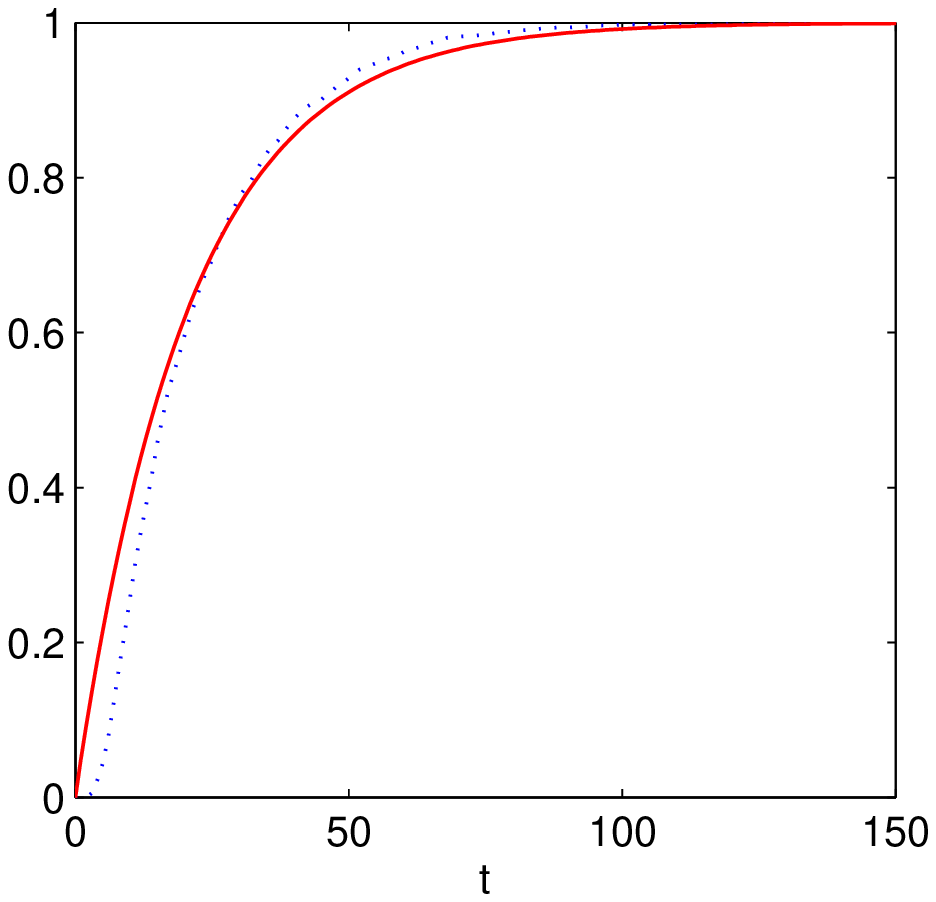}
        }
        \caption{Length between two consecutive up-crossings of a large negative level $-\gamma$, evaluated with squared exponential ACF with $\dc = 2$ and $-\gamma = -1$. (a) Plot of the histogram. (b) Plot of $F_1(t)$ against normalized cumulated sum.}
        \label{Fig:lowConUcr}
    \end{figure}

    In particular, for $k = 1$ we obtain the distribution of the time between two consecutive up-crossings:
    \begin{equation}\label{Eq:Dist2Up-crossings}
        \lim_{\gamma \to \infty}F_{1}\left(\frac{t}{\mu}\right) = 1 - e^{-t}.
    \end{equation}
    \fref{Fig:lowConUcr} confirms the effectiveness of this exponential property by simulation of a standard normal process with the squared exponential autocorrelation function \eref{eq:acfdblexp}.

\section{Excursions Above a Large Negative Level}\label{Sec:ExcurSmallLevel}

    Given a large negative level $-\gamma \to -\infty$, we are interested in the interval $\tauu$ of an up-excursion of the process $X(t)$ above $-\gamma$. We investigate its limiting distribution by using Theorem~\ref{Thm:IntervalBetweenUpcrosses}. For this, assume that the stationary normal process $X(t)$ admits an autocorrelation function $R_X(\tau)$ satisfying conditions \eqref{Eq:ACF_Cond4} and \eqref{Eq:ACF_Cond5}.

    Let $z$ be the interval between a down-crossing to the next subsequent down-crossing of level $-\gamma$. By the aforementioned symmetry property, the distribution of $z$ as $-\gamma \to -\infty$ is identical to that of the time between an up-crossing and the next subsequent up-crossing of the level $\gamma \to \infty$. Thus, under the conditions stated above, the limiting distribution of $z$ as $-\gamma \to -\infty$ is given according to Theorem~\ref{Thm:IntervalBetweenUpcrosses} for $k = 1$:
    \begin{equation}\label{Eq:DistTau2}
        F_{Z}(z) = 1 - e^{-\mu z}, \quad \textrm{as } -\gamma \to -\infty
    \end{equation}
    with $\mu = \E D_{-\gamma}$, which is the down-crossing rate of level $-\gamma$. By Proposition~\ref{Prop:UpDownCrossRate}, we can write $\mu = \E U_{-\gamma}$.

    Let $\taud$ be the time between a down-crossing and the next up-crossing of the level $-\gamma$. Then, the interval $\tauu$ of an up-excursion above $-\gamma$ is precisely the interval from the up-crossing to the next down-crossing of $X(t)$ of the level $-\gamma$. It is clear that $\tauu$ is a random variable given as
    $$\tauu = z - \taud \, | \, z \geq \taud.$$
    But $z$ is an exponential distribution with rate $\mu$ as given by \eqref{Eq:DistTau2}. By the memorylessness property of the exponential distribution, $\tauu$ is an exponential distribution with rate $\mu$. This is stated in the following.
    \begin{theorem}\label{Thm:AsympExcurSmall}
        With the process $X(t)$ described above, under the conditions \eqref{Eq:ACF_Cond4} and \eqref{Eq:ACF_Cond5}, the time $\tauu$ of an up-excursion of $X(t)$ above a large negative level $-\gamma < 0$ asymptotically follows an exponential distribution of rate $\mu = \E U_{-\gamma}$, i.e.,
        $$\Pb(\tauu \leq \tau) = 1 - e^{-\mu \tau},  \quad \textrm{as } -\gamma \to -\infty.$$
    \end{theorem}

   \begin{figure}
        \centering
        \subfigure[Histogram of $\tauu$]
        {
            \includegraphics[width=0.45\textwidth]{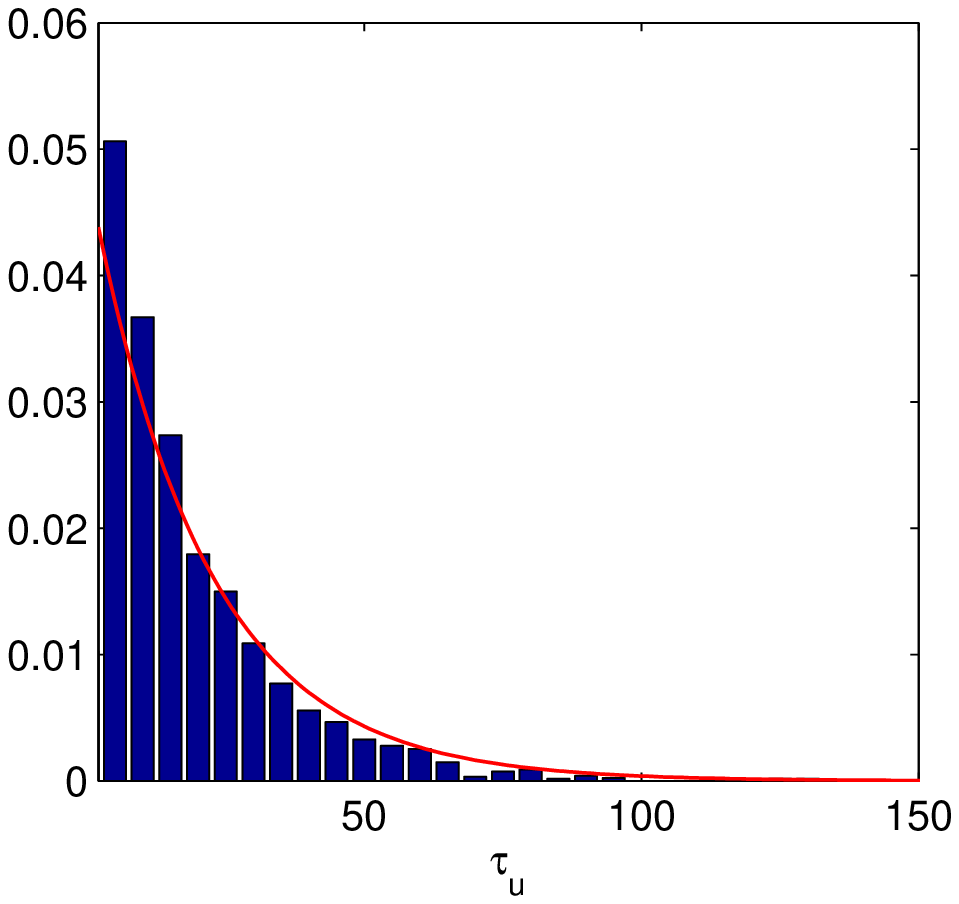}
        }
        \subfigure[$\Pb(\tauu \leq \tau)$]
        {
            \includegraphics[width=0.45\textwidth]{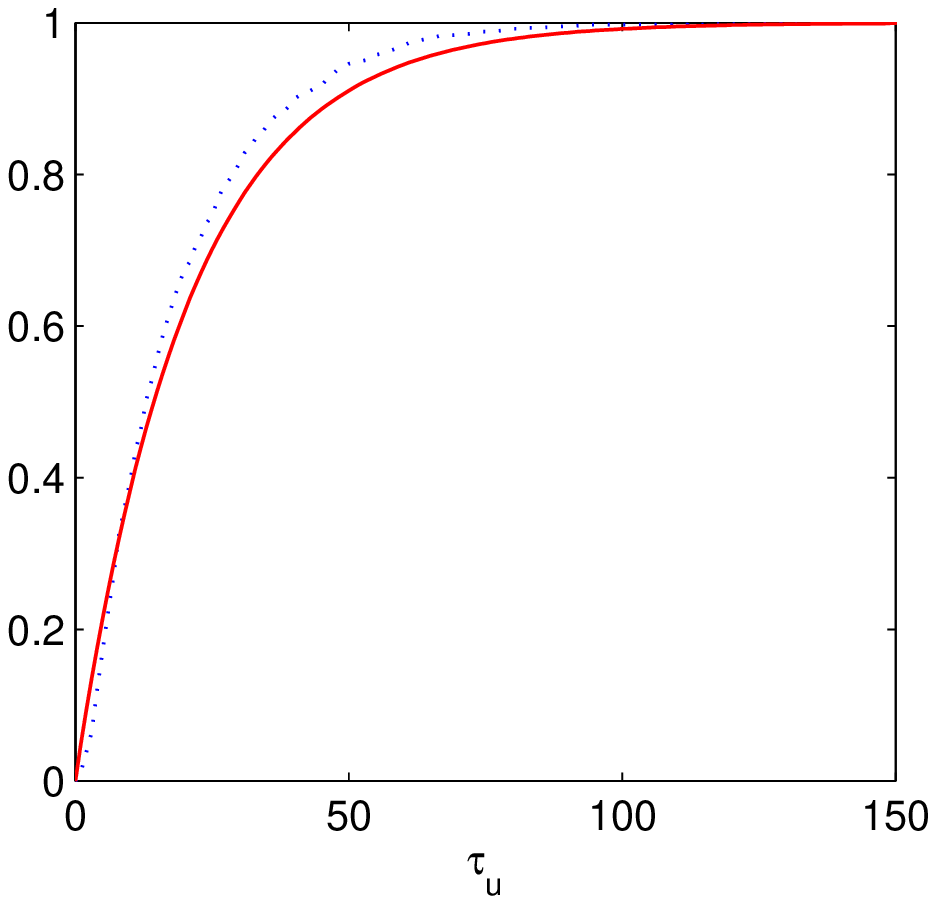}
        }
        \caption{Distribution of large negative excursions evaluated with squared exponential ACF, $\dc = 2$, and $-\gamma = -1$. (a) Plot of the asymptotic density (solid) against the histogram. (b) Asymptotic distribution (solid) against the normalized cumulated sum.}
        \label{Fig:lowExcur}
    \end{figure}

    By Proposition~\ref{Prop:UpDownCrossRate} and Theorem~\ref{Thm:LCR} we have that
    $$\mu = \E U_{-\gamma} = \frac{1}{2\pi}\sqrt{\frac{\lambda_2}{\lambda_0}}\exp\left(-\frac{\gamma^2}{2\lambda_0}\right),$$
    which tends to zero as $-\gamma \to -\infty$. As a consequence, $\tauu \to \infty$ asymptotically almost surely (a.a.s.) as $-\gamma \to -\infty$. Whereas, by Theorem~\ref{Thm:AsympExcursion}, the length of an up-excursion above a very large level $\gamma$ behaves asymptotically as a Rayleigh distribution with parameter $2/(\gamma\sqrt{\lambda_2})$, which tends to zero as $\gamma \to \infty$. Thus, the length of large excursions converges a.a.s. to zero. This means that the process $X(t)$ stays most of the time above a large negative level, whereas it stays above a very large level only during a very short interval.

    The above difference leads to another distinction between up-excursions above a large level $\gamma$  (\emph{large excursions}) and those above a large negative level $-\gamma$ (\emph{large negative excursions}). In the former case, the length of a large excursion is very small so that their trajectory can be predicted by using the information provided by the autocorrelation function, as shown by Theorem~\ref{Thm:AsympExcursion}. By contrast, the length of a large negative excursion is very long as discussed above, and it is not possible to predict $X(t_0 + T)$ from $X(t_0)$ under the condition \eqref{Eq:ACF_Cond5} when $T \to \infty$. Therefore, we do not have precise information about the trajectory of large negative excursions except for their length, given according to Theorem~\ref{Thm:AsympExcurSmall}.

    \fref{Fig:lowExcur} compares the analytical results against the simulation results on the length of large negative excursions. We can see that the asymptotic property obtained using Theorem~\ref{Thm:AsympExcurSmall} holds well even with the relatively small magnitude $-\gamma$.

\section{Crossings of Successive Large Levels}\label{Sec:ExcurMultLevel}

    Until now we have been concerned with crossings of a stationary Gaussian process $X(t)$ with respect to one level. In this section, we investigate crossings of $X(t)$ of two successive large levels by using the asymptotic trajectory of large excursions provided by Theorem~\ref{Thm:AsympExcursion}. We assume that the autocorrelation function $R_X(\tau)$ of $X(t)$ satisfies conditions \eqref{Eq:ACF_Cond2} and \eqref{Eq:ACF_Cond3} so that Theorem~\ref{Thm:AsympExcursion} applies. Note that here we relax the assumption \eqref{Eq:ACF_Cond4} used in \sref{Sec:ExcurSmallLevel} that $R_X(\tau)$ has a finite fourth derivative at the origin. For the following development, let us make the following definition:
    \begin{itemize}
        \item $\exu(\gamma)$: the event that $X(t)$ has an up-excursion above level $\gamma$,
        \item $\exd(\gamma)$: the event that $X(t)$ has a down-excursion below level $\gamma$.
    \end{itemize}
    We also expand the above definition by adding a time duration:
    \begin{itemize}
        \item $\exu(\gamma, \tau)$: the event that $X(t)$ has an up-excursion of length $\tau$ above level $\gamma$,
        \item $\exd(\gamma, \tau)$: the event that $X(t)$ has a down-excursion of length $\tau$ below level $\gamma$.
    \end{itemize}

    \subsection{Mean Number of Crossings of Successive Large Levels}

    Given that $X(t)$ has an up-excursion above a large level $\gamma_1$ with length $T_1 \geq \tau_1$ for some $\tau_1$, we investigate the mean number of crossings of $X(t)$ of a level $\gamma_2 \geq \gamma_1$, and in particular, we study this quantity as $\gamma_1 \to +\infty$.

    Under conditions \eqref{Eq:ACF_Cond2} and \eqref{Eq:ACF_Cond3}, Theorem~\ref{Thm:AsympExcursion} states that an excursion of $X(t)$ above level $\gamma_1$ behaves asymptotically as
    \begin{equation*}
        X(t) \sim \gamma_1 + \xi t - \gamma_1 \frac{\lambda_2 t^2}{2}, \quad \textrm{as } \gamma_1 \to +\infty,
    \end{equation*}
    where $\xi$ is a Rayleigh random variable with parameter $\sqrt{\lambda_2}$. This suggests that the length of an excursion of $X(t)$ above level $\gamma_1$ is
    \begin{equation}\label{Eq:Texcur1}
        T_1 =  \frac{2}{\gamma_1\lambda_2} \xi,
    \end{equation}
    and during which $X(t)$ behaves as a downwards parabola. During this time interval $T_1$, $X(t)$ will have one up-crossing of level $\gamma_2$ if
    \begin{equation}\label{Eq:ExcurGam2}
        \gamma_1 + \xi t - \gamma_1 \frac{\lambda_2 t^2}{2} > \gamma_2,
    \end{equation}
    and will have zero up-crossing of level $\gamma_2$ otherwise. Here, $X(t)$ will be tangent to the level $\gamma_2$ if the discriminant $\Delta = \xi^2 - 2\gamma_1\lambda_2(\gamma_2-\gamma_1)$ is equal to 0, and this is not considered as an up-crossing. Solving for the quadratic inequality $\Delta > 0$, we obtain
    \begin{equation*}
        \textrm{\# up-crossings of } X(t) \textrm{ above } \gamma_2 \, | \, \exu(\gamma_1) = \begin{cases} 1, & \Delta > 0 \\ 0, & \Delta \leq 0 \end{cases}.
    \end{equation*}
    Hence, the mean number of up-crossings of level $\gamma_2 \geq \gamma_1$ during duration $T_1$ given that $X(t)$ has an up-excursion above $\gamma_1$ with length $T_1 \geq \tau_1$ is
    \begin{align*}
        \E\{\textrm{\# up-crossings of } X(t) \textrm{ above } \gamma_2 \, | & \, \exu(\gamma_1, T_1) \textrm{ with } T_1 \geq \tau_1\} \\
        & = \Pb(\Delta > 0 \, | \, T_1 \geq \tau_1) \\
        & = \Pb\{\xi^2 > 2\gamma_1\lambda_2(\gamma_2 - \gamma_1) \, | \, \xi \geq \frac{\gamma_1\lambda_2}{2}\tau_1\} \\
        & = \frac{\Pb\left\{\xi^2 > \max\left[2\gamma_1\lambda_2(\gamma_2 - \gamma_1), (\frac{\gamma_1\lambda_2}{2}\tau_1)^2\right]\right\}}{\Pb\left(\xi \geq \frac{\gamma_1\lambda_2}{2}\tau_1\right)},
    \end{align*}
    as $\gamma_1 \to +\infty$. With $\xi$ a Rayleigh distribution of parameter $\sqrt{\lambda_2}$, $\xi^2$ is exponentially distributed with intensity $1/(2\lambda_2)$:
    $$\Pb(\xi^2 \leq x) = 1 - \exp\left(-\frac{x}{2\lambda_2}\right).$$
    Hence,
    \begin{align*}
        \Pb\left\{\xi^2 > \max[2\gamma_1\lambda_2(\gamma_2 - \gamma_1), (\frac{\gamma_1\lambda_2}{2}\tau_1)^2]\right\}
        & = \exp\left\{- \max[\gamma_1(\gamma_2 - \gamma_1), \frac{\gamma_1^2\lambda_2}{8}\tau_1^2]\right\}\\
        & = \exp\{- V \max[(\tau_1^{\ast})^2, \tau_1^2]\}
    \end{align*}
    where
    \begin{equation}\label{Eq:V}
        V := \frac{\gamma_1^2\lambda_2}{8},
    \end{equation}
    and
    \begin{equation}\label{Eq:Texcur1_CritialPoint}
        \tau_1^{\ast} := \sqrt{\frac{\gamma_1(\gamma_2 - \gamma_1)}{V}} = \sqrt{\frac{8(\gamma_2 - \gamma_1)}{\gamma_1\lambda_2}},
    \end{equation}

    The above analysis leads to the following conclusion
    \begin{theorem}\label{Thm:UpCrossCondGen}
        With the process $X(t)$ described above, assume that $R_X(t)$ satisfies conditions \eqref{Eq:ACF_Cond2} and \eqref{Eq:ACF_Cond3}. Then, for $\gamma_2 \geq \gamma_1$
        \begin{equation*}
        \E\{\textrm{\# up-crossings of } X(t) \textrm{ of } \gamma_2 \, | \, \exu(\gamma_1, T_1) \textrm{ with } T_1 \geq \tau_1\} = \frac{\exp\left\{- V \max[(\tau_1^{\ast})^2, \tau_1^2]\right\}}{\exp\left(-V \tau_1^2\right)},
        \end{equation*}
        as $\gamma_1 \to +\infty$ with $V$ given in \eqref{Eq:V} and $\tau_1^{\ast}$ given in \eqref{Eq:Texcur1_CritialPoint}.
    \end{theorem}

    In particular, if the up-excursion of $X(t)$ above $\gamma_1$ is long enough so that $\tau_1 > \tau_1^{\ast}$, then there will be one up-crossing of $X(t)$ above the level $\gamma_2$ with probability 1. Motivated by this, we call $\tau_1^{\ast}$ the \emph{critical length} of an up-excursion above $\gamma_1$.

    A particular application of the above result is for the case of $\tau_1 = 0$. Denote by $U_{\gamma_2 | \gamma_1}$ the number of up-crossings of $X(t)$ above level $\gamma_2$ given that $X(t)$ is above $\gamma_1$. Then $U_{\gamma_2 | \gamma_1}$ is given as
    \begin{equation*}
    U_{\gamma_2 | \gamma_1} = \E\{\textrm{\# up-crossings of } X(t) \textrm{ above } \gamma_2 \, | \, \exu(\gamma_1, T_1) \textrm{ with } T_1 \geq 0\}.
    \end{equation*}
    Hence, we have:
    \begin{corollary}\label{Cor:Thm:UpCrossCondGen}
        With the notation described above and hypothesis of Theorem~\ref{Thm:UpCrossCondGen}:
        \begin{equation}\label{Eq:UpCrossCond}
        \E U_{\gamma_2 | \gamma_1} = \exp\{-\gamma_1(\gamma_2 - \gamma_1)\} \quad \textrm{as } \gamma_1 \to +\infty.
        \end{equation}
    \end{corollary}

    During the up-excursion of $X(t)$ above $\gamma_1$, if $X(t)$ has an up-crossing of $\gamma_2$, then $X(t)$ will have a down-crossing of $\gamma_2$ by the parabola property of $X(t)$ as $\gamma_1 \to +\infty$. Hence, denoting $D_{\gamma_2 | \gamma_1}$ to be the number of down-crossings of $X(t)$ below the level $\gamma_2$ given that $X(t)$ has an up-excursion above $\gamma_1$, it is obvious that
    $$\E D_{\gamma_2 | \gamma_1} =  \E U_{\gamma_2 | \gamma_1},$$
    and the mean number of crossings is
    $$\E C_{\gamma_2 | \gamma_1} =  2 \, \E U_{\gamma_2 | \gamma_1},$$
    as $\gamma_1 \to +\infty$.


    Above, we obtained the mean number of up-crossings above (and of down-crossings below) a subsequent level $\gamma_2$ given an up-excursion above a lower level $\gamma_1 \to +\infty$. Using these results, we can also obtain the mean up-crossing rate (and also mean down-crossing rate) of $[(X(t) \, | \, \exu(\gamma_1, T_1) \textrm{ with }  T_1 \geq \tau_1)]$ above (below) the level $\gamma_2 \geq \gamma_1$ as $\gamma_1 \to +\infty$ by noting that the up-crossing rate is given as the ratio of number of up-crossings divided by the time duration $T_1$. Taking the expectation of this ratio for $T_1$ from $\tau_1$ to infinity, we obtain the mean up-crossing rate without any difficulty. However, we prefer to not provide it here since such a mean up-crossing rate does not provide meaningful information, and may even introduce ambiguity. In fact, a mean crossing rate should give the mean number of crossings when it is multiplied by a time duration, whereas in our case we know that $[(X(t) \, | \, \exu(\gamma_1))]$ can have at most only one up-crossing of $\gamma_2$ for all time durations $T_1$.

   \begin{figure}
        \centering
        \subfigure
        {
            \includegraphics[width=0.45\textwidth]{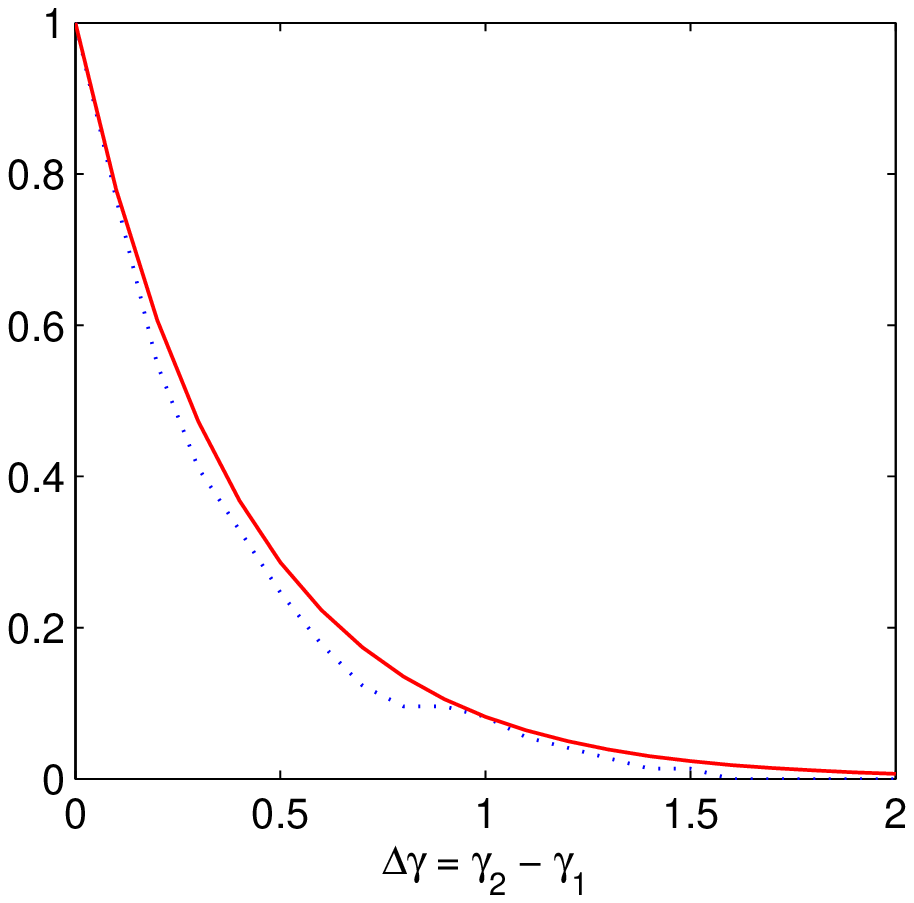}
        }
        \subfigure
        {
            \includegraphics[width=0.45\textwidth]{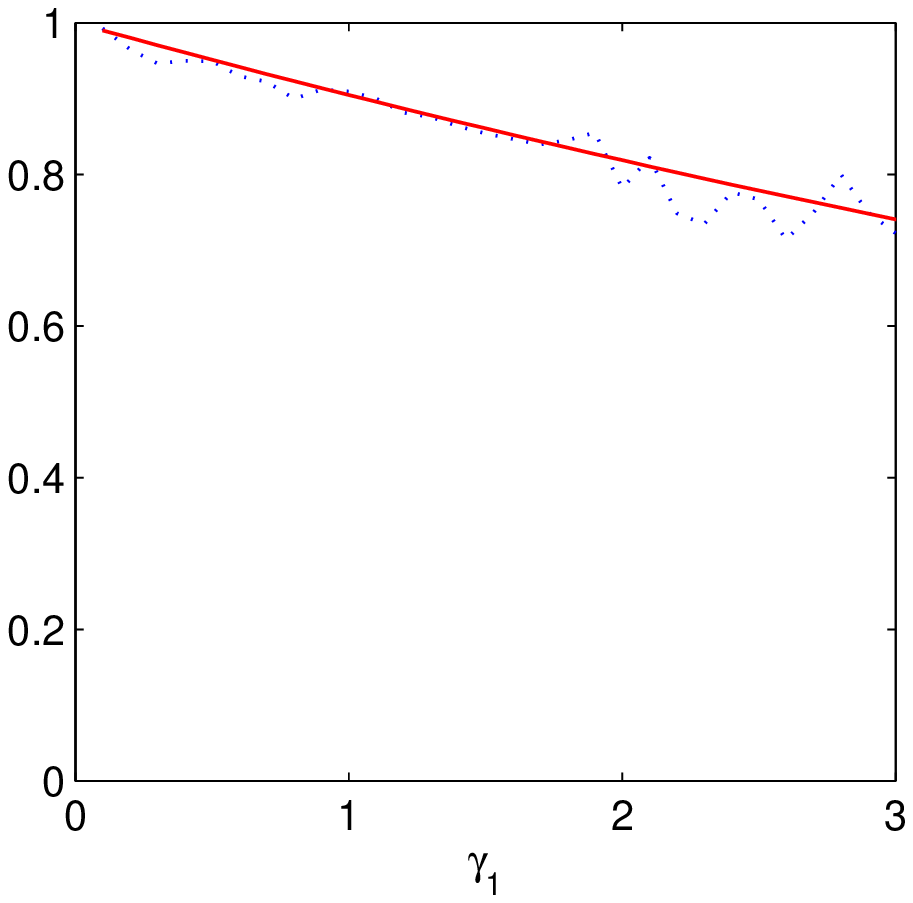}
        }
        \caption{Plots of $\E U_{\gamma_2|\gamma_1}$ evaluated with squared exponential ACF with $\dc = 2$. In both figures, solid lines correspond to analytical forms, and dotted lines correspond to simulation. (a) Versus level difference $\Delta\gamma = \gamma_2 - \gamma_1$ with $\gamma_1 = 2.5$. (b) Versus different $\gamma$ given fixed $\Delta\gamma = 0.1$.}\label{Fig:meanSucUcr}
    \end{figure}

    \fref{Fig:meanSucUcr} checks the effectiveness of the asymptotic results obtained above against simulation. We see that the asymptotic result of $\E U_{\gamma_2|\gamma_1}$ as given by \eref{Eq:UpCrossCond} holds well even over practical ranges of $\gamma_1$ and $\gamma_2$. In \fref{Fig:meanSucUcr}(b), increasing fluctuations of the simulation curve (dotted line) are observed at the right end, which is due to the fact that larger values of $\gamma_1$ reduce the number of up-excursions above $\gamma_1$, leading to fewer simulation samples in estimating the statistics.

\subsection{Length of Excursions Above Successive Levels}

    Using the notation described above, we devote to this subsection the distribution of the length, say $T_2$, of an up-excursion of $X(t)$ above the level $\gamma_2 \geq \gamma_1$ given that $X(t)$ has an up-excursion above $\gamma_1$ with length $T_1 \geq \tau_1$, as $\gamma_1 \to +\infty$.  The conditional length $T_2$ is the length of the time interval during which the inequality \eqref{Eq:ExcurGam2} holds:
    \begin{equation*}
    [T_2  \, | \, \exu(\gamma_1, T_1) \textrm{ with } T_1 \geq \tau_1] = \begin{cases}\frac{2\sqrt{\Delta}}{\gamma_1\lambda_2}, & \Delta > 0 \\ 0, & \Delta \leq 0 \end{cases},
    \end{equation*}
    where recall that $\Delta$ is the discriminant. Hence,
    \begin{align*}
        \Pb\{T_2 = 0 \, | \, \exu(\gamma_1, T_1) \textrm{ with } T_1 \geq \tau_1\}
        & = \Pb\{\Delta \leq 0 \, | \, T_1 \geq \tau_1\}\\
        & = \Pb\Big\{\xi^2 \leq 2\gamma_1\lambda_2(\gamma_2-\gamma_1) \, \big| \, \xi \geq \frac{\gamma_1\lambda_2}{2}\tau_1\Big\}\\
        & = \frac{\Pb\Big\{(\frac{\gamma_1\lambda_2}{2}\tau_1)^2 \leq \xi^2 \leq 2\gamma_1\lambda_2(\gamma_2-\gamma_1)\Big\}}{\Pb(\xi \geq \frac{\gamma_1\lambda_2}{2}\tau_1)},
    \end{align*}
    which, by the Rayleigh distribution of $\xi$, reduces to
    \begin{equation}\label{Eq:DistT2_1}
        \Pb\{T_2 = 0 \, | \, \exu(\gamma_1, T_1) \textrm{ with } T_1 \geq \tau_1\} = \one\left(\tau_1 \leq \tau_1^{\ast}\right)\Big[1 - \frac{e^{-V(\tau_1^{\ast})^2}}{ e^{-V \tau_1^2}}\Big]
    \end{equation}
    with $V$ given by Theorem~\ref{Thm:UpCrossCondGen}, and where $\tau_1^{\ast}$ is the critical length given by \eqref{Eq:Texcur1_CritialPoint}. Similarly, for all $\tau \geq 0$
    \begin{align*}
        \Pb\{T_2 > \tau \, | \, \exu(\gamma_1, T_1) \textrm{ with } T_1 \geq \tau_1\} & = \Pb\Big\{\frac{2\sqrt{\Delta}}{\gamma_1\lambda_2} > \tau \, \big| \, T_1 \geq \tau_1\Big\} \\
        & = \Pb\Big\{\xi^2 > (\frac{\gamma_1\lambda_2}{2}\tau)^2 + 2\gamma_1\lambda_2(\gamma_2-\gamma_1) \, \Big| \, \xi \geq \frac{\gamma_1\lambda_2}{2}\tau_1\Big\}\\
        & = \frac{\Pb\Big\{\xi^2 > \max[(\frac{\gamma_1\lambda_2}{2}\tau)^2 + 2\gamma_1\lambda_2(\gamma_2-\gamma_1), (\frac{\gamma_1\lambda_2}{2}\tau_1)^2]\Big\}}{\Pb(\xi \geq \frac{\gamma_1\lambda_2}{2}\tau_1)},
    \end{align*}
    which is reduces to
    \begin{equation}\label{Eq:DistT2_2}
        \Pb\{T_2 > \tau \, | \, \exu(\gamma_1, T_1) \textrm{ with } T_1 \geq \tau_1\} = \frac{\exp\Big\{-V \max[\tau^2 + (\tau_1^{\ast})^2, \tau_1^2]\Big\}}{\exp(-V\tau_1^2)}.
    \end{equation}

    We conclude the above results in the following
    \begin{proposition}\label{Prop:ExcurTimeCond}
        With the notation described above and the conditions of Theorem~\ref{Thm:UpCrossCondGen}, the distribution of the length $T_2$ of an up-excursion of $X(t)$ above $\gamma_2$ given that $X(t)$ has an up-excursion above $\gamma_1 \leq \gamma_2$ with length $T_1 \geq \tau_1$ is determined jointly by \eqref{Eq:DistT2_1} and \eqref{Eq:DistT2_2} as $\gamma_1 \to +\infty$.
    \end{proposition}

   \begin{figure}
        \centering
        \subfigure
        {
            \includegraphics[width=0.45\textwidth]{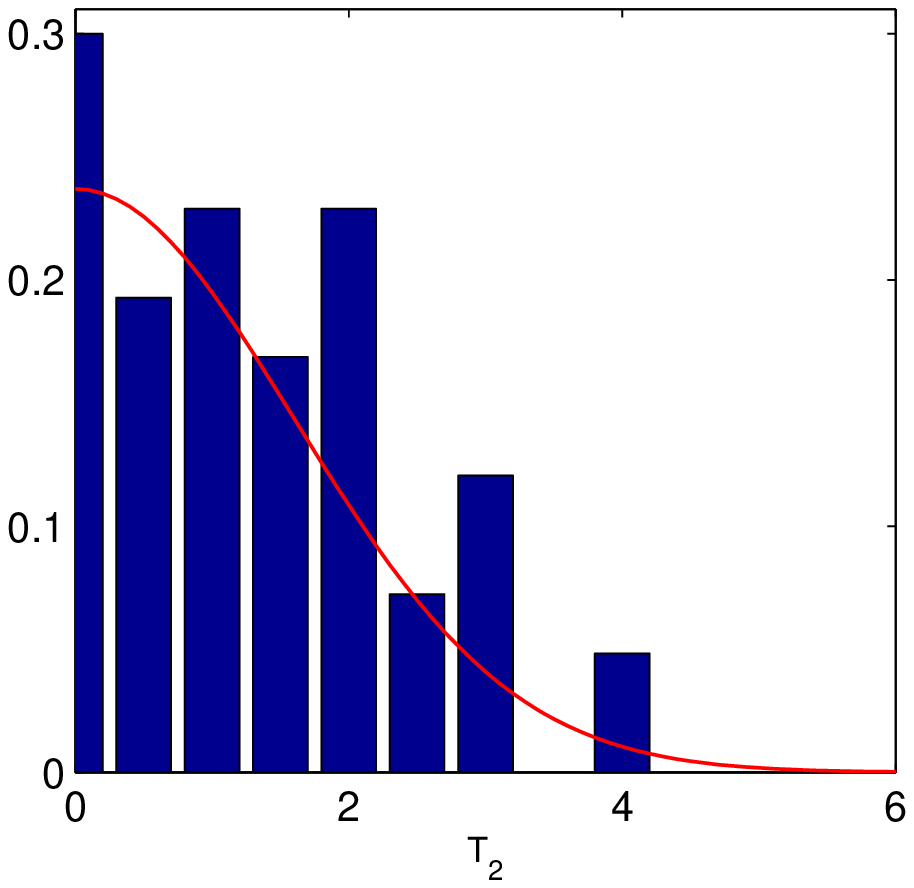}
        }
        \subfigure
        {
            \includegraphics[width=0.45\textwidth]{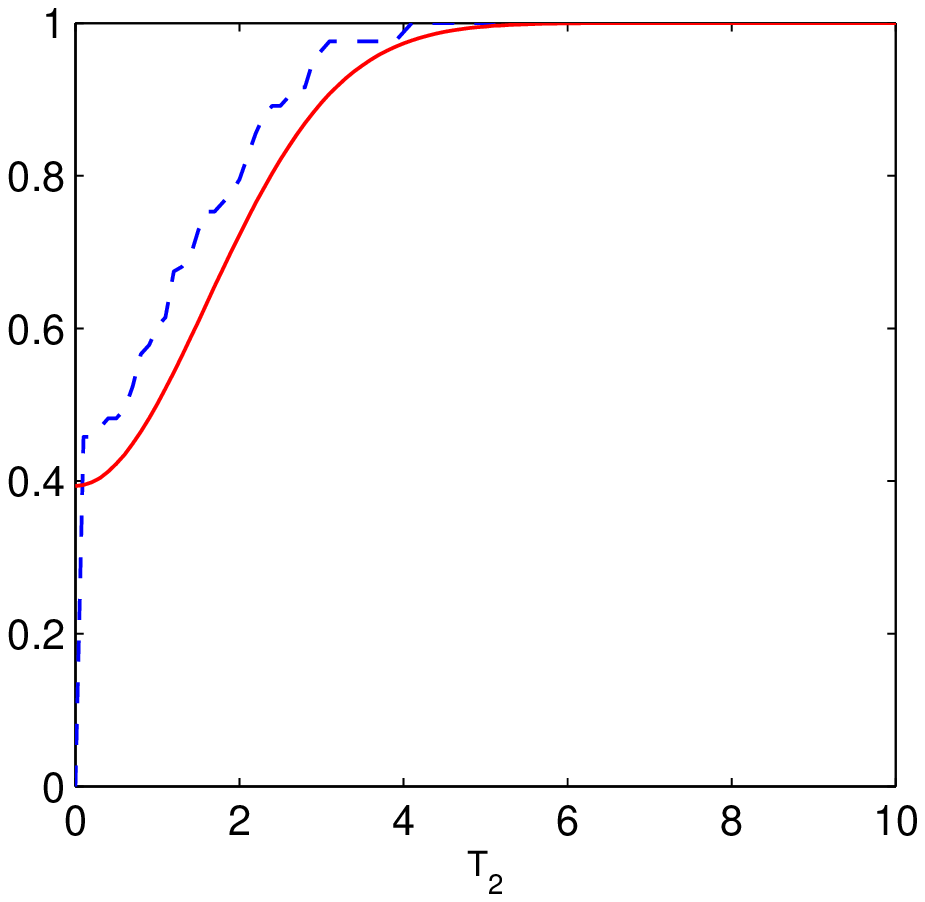}
        }
        \caption{Distribution of the length of successive up-excursions above level $\gamma_2 \geq \gamma_1$ given up-excursion above large level $\gamma_1$, evaluated with squared exponential ACF, $\gamma_1 = 2.5$, and $\gamma_2 = \gamma_1+0.2$. The solid lines correspond to analytical results, and histogram and dashed line correspond to the simulation.}\label{Fig:lenSucUcr}
    \end{figure}

    According to \eref{Eq:DistT2_1}, $T_2$ has a mass at the origin. This is mainly contributed by excursions above $\gamma_1$ whose down-ward parabola peak value is not critically high enough to cross the successive higher level $\gamma_2$. \fref{Fig:lenSucUcr} shows that the distribution of the length of successive excursions provided by Proposition~\ref{Prop:ExcurTimeCond} effectively matches with the simulation.

\section{Application}\label{Ssec:ExcurMultLevelApp}

    Wireless communications quality depends closely on the received signal. Especially, the service may be interrupted when the received signal strength is lower than some critical level so that useful symbols are severely corrupted by the background noise and interference. It is well known that radio transmission is sensitive to the propagation condition which is subject to random variation behaving like a stochastic process. The problem has therefore been to understand and characterize its properties, in particular, in terms of amplitude fluctuations with respect to some thresholds.

    Basically, a radio link failure occurs when the signal power falls below a threshold and stays below during some minimum duration. This is a critical problem of all radio communication systems, and have been analyzed using the asymptotic properties of large excursions of a stationary Gaussian process, e.g., \cite{Mandayam1998,Nguyen2011Icc}. Recent mobile systems, such as 3GPP 3G and 4G systems, introduce some kind of preventive measure to ameliorate the above mentioned radio link failure problem by searching for a new base station when the signal power of the tagged base station stays below a threshold over a predefined time duration. Obviously, this threshold should be larger than the link failure level. As such, knowing excursion properties of successive levels would be helpful to analyze this behavior. The following application of the developed theory addresses this technical problem, and of course can be used for any similar problem.

    Using the notation and conditions described in \sref{Sec:ExcurMultLevel}, for two fixed levels $\gamma_1$ and $\gamma_2 \geq \gamma_1$, we are interested in successive up-excursions of $X(t)$ above $\gamma_1$ and above $\gamma_2$ during some time window $[0, T]$. Specifically, we determine the following probability
    \begin{align*}
    P & := \Pb\{\exu(\gamma_1, T_1) \textrm{ with } T_1 \geq \tau_1 \textrm{, and } \exu(\gamma_2, T_2) \textrm{ with } T_2 \geq \tau_2 \textrm{ for } t \in [0, T]\} \\
    & = \frac{\E\{\textrm{total time of } \exu(\gamma_1, T_1) \textrm{ with } T_1 \geq \tau_1 \textrm{, and } \exu(\gamma_2, T_2) \textrm{ with } T_2 \geq \tau_2, \, t \in [0, T]\}}{T},
    \end{align*}
    where we have made the ergodicity assumption of $X(t)$. Of course, the main concern here is due to the numerator.
    Putting
    $$T_2^{\ast} = T_2 \, | \, \exu(\gamma_1, T_1) \textrm{ with } T_1 \geq \tau_1,$$
    where, according to the notation described above, $T_2$ is the length of an up-excursion of $X(t)$ above $\gamma_2$. Thus, we have
    \begin{align}
        \E\{\textrm{total time of } & \exu(\gamma_1, T_1) \textrm{ with } T_1 \geq \tau_1 \textrm{, and } \exu(\gamma_2, T_2) \textrm{ with } T_2 \geq \tau_2, \, t \in [0, T]\} \nonumber\\
        = & \E\{\textrm{\# excursions above } \gamma_1 \textrm{ with } T_1 \geq \tau_1 \textrm{ for } t \in [0, T]\} \nonumber\\
        \times & \E\{\textrm{\# excursions above } \gamma_2 \textrm{ with } T_2 \geq \tau_2 \, | \, X(t) \geq \gamma_1 \textrm{ with } T_1 \geq \tau_1\}\nonumber\\
        \times & \E\{T_2^{\ast} \, | \, T_2^{\ast} \geq \tau_2\} \label{Eq:DblExcur_0}.
    \end{align}
    We shall determine the three terms on the right hand side of \eref{Eq:DblExcur_0} in the following. Firstly we have
    \begin{align}
        \E\{\textrm{\# excursions above } \gamma_1 & \textrm{ with } T_1 \geq \tau_1 \textrm{ for } t \in [0, T]\} \nonumber\\
        & = \E\{\textrm{\# up-crossings of } \gamma_1 \textrm{ during } [0, T]\} \times \Pb(T_1 \geq \tau_1) \nonumber\\
        & = T \cdot \E U_{\gamma_1} \cdot \exp(-V\tau_1^2), \quad \textrm{as }\gamma_1 \to +\infty, \label{Eq:DblExcur_Part1}
    \end{align}
    which is obtainable with $T_1$ given by \eqref{Eq:Texcur1}, where $\E U_{\gamma_1}$ is the mean up-crossing rate of level $\gamma_1$ given according to Proposition~\ref{Prop:UpDownCrossRate}.

    For the second term on the right-hand side of \eqref{Eq:DblExcur_0},
    \begin{align*}
    \E\{\textrm{\# excursions above } \gamma_2 & \textrm{ with } T_2 \geq \tau_2 \, | \, \exu(\gamma_1, T_1) \textrm{ with } T_1 \geq \tau_1\} \\
    = & \E\{\textrm{\# up-crossings of } \gamma_2 \, | \, \exu(\gamma_1, T_1) \textrm{ with } T_1 \geq \tau_1\} \\
    & \times \Pb\{T_2 \geq \tau_2 \, | \, \exu(\gamma_1, T_1) \textrm{ with } T_1 \geq \tau_1\},
    \end{align*}
    in which the first term on the right hand side is the mean number of up-crossings of $X(t)$ above level $\gamma_2$ given that $X(t)$ has an up-excursion above $\gamma_1$ of length $T_1 \geq \tau_1$, which is given according to Theorem~\ref{Thm:UpCrossCondGen}; and the second term on the right-hand side is given as
    \begin{align}
    \Pb\{&T_2 \geq \tau_2 \, | \, \exu(\gamma_1, T_1) \textrm{ with } T_1 \geq \tau_1\} \nonumber\\
    & =  \Pb\{T_2 > \tau_2 \, | \, \exu(\gamma_1, T_1) \textrm{ with } T_1 \geq \tau_1\} + \Pb\{T_2 = \tau_2 \, | \, \exu(\gamma_1, T_1) \textrm{ with } T_1 \geq \tau_1\}\nonumber\\
    & = \frac{e^{-V \max\{\tau_2^2 + (\tau_1^{\ast})^2, \tau_1^2\}}}{e^{-V\tau_1^2}} + \one(\tau_2 =0)\one(\tau_1 \leq \tau_1^{\ast})\Big[1 - \frac{e^{-V (\tau_1^{\ast})^2}}{ e^{-V \tau_1^2}}\Big]\label{Eq:DistT2}
    \end{align}
    which is obtainable by Proposition~\ref{Prop:ExcurTimeCond}. Therefore,
    \begin{multline}
        \E\{\textrm{\# excursions above } \gamma_2 \textrm{ with } T_2 \geq \tau_2 \, | \, \exu(\gamma_1, T_1) \textrm{ with } T_1 \geq \tau_1\} \\
        = \frac{e^{-V \max\{(\tau_1^{\ast})^2, \tau_1^2\}}}{e^{-V \tau_1^2}} \Bigg\{\frac{e^{-V \max\{\tau_2^2 + (\tau_1^{\ast})^2, \tau_1^2\}}}{e^{-V\tau_1^2}} + \one(\tau_2 =0)\one(\tau_1 \leq \tau_1^{\ast})\Big[1 - \frac{e^{-V (\tau_1^{\ast})^2}}{ e^{-V \tau_1^2}}\Big]\Bigg\} \label{Eq:DblExcur_Part2}
    \end{multline}

    Finally, we derive the last term on the right-hand side of \eqref{Eq:DblExcur_0}.
    \begin{equation}\label{Eq:DblExcur_Part3}
        \E\{T_2^{\ast} \, | \, T_2^{\ast} \geq \tau_2\} = \tau_2 + \int_{\tau_2}^{\infty}\frac{\Pb(T_2^{\ast} > \tau)}{\Pb(T_2^{\ast} \geq \tau_2)}\di\tau
    \end{equation}
    where by \eqref{Eq:DistT2_2} we have
    \begin{equation*}
        \int_{\tau_2}^{\infty}\Pb(T_2^{\ast} > \tau)\di\tau  = \int_{\tau_2}^{\infty} \frac{e^{-V \max\{\tau^2 + (\tau_1^{\ast})^2, \tau_1^2\}}}{e^{-V\tau_1^2}}\di\tau.
    \end{equation*}
    For this we distinguish between two cases with $\tau_1 \leq \tau_1^{\ast}$ and $\tau_1 > \tau_1^{\ast}$.
    (i) For $\tau_1 \leq \tau_1^{\ast}$,
    \begin{equation}\label{Eq:DblExcur_Part3_i}
    \int_{\tau_2}^{\infty}\Pb(T_2^{\ast} > \tau)\di\tau  = \int_{\tau_2}^{\infty} \frac{e^{-V(\tau^2 + (\tau_1^{\ast})^2)}}{e^{-V\tau_1^2}}\di\tau = \frac{e^{-V (\tau_1^{\ast})^2}}{e^{-V\tau_1^2}} \sqrt{\frac{\pi}{4V}}\erfc(\sqrt{V}\tau_2).
    \end{equation}
    (ii) for $\tau_1 > \tau_1^{\ast}$, let us denote by $\tau_2^{\ast}$ the solution of
    $\tau^2 + (\tau_1^{\ast})^2 = \tau_1^2$, given by
    \begin{equation}\label{Eq:Texcur2_CriticalPoint}
        \tau_2^{\ast} = \sqrt{\tau_1^2 - (\tau_1^{\ast})^2}.
    \end{equation}
    Then,
    \begin{align}
    \int_{\tau_2}^{\infty}\Pb & (T_2^{\ast} > \tau)\di\tau  = \one(\tau_2 < \tau_2^{\ast})\int_{\tau_2}^{\tau_2^{\ast}}\di\tau + \int_{\max\{\tau_2^{\ast}, \tau_2\}}^{\infty} \frac{e^{-V(\tau^2 + (\tau_1^{\ast})^2)}}{e^{-V\tau_1^2}}\di\tau \nonumber\\
    & = \one(\tau_2 < \tau_2^{\ast})(\tau_2^{\ast} - \tau_2) + \frac{e^{-V(\tau_1^{\ast})^2}}{e^{-V\tau_1^2}} \sqrt{\frac{\pi}{4V}}\erfc[\sqrt{V}\max\{\tau_2^{\ast}, \tau_2\}].\label{Eq:DblExcur_Part3_ii}
    \end{align}
    So, $\E\{T_2^{\ast} \, | \, T_2^{\ast} \geq \tau_2\}$ is given according to \eqref{Eq:DblExcur_Part3} with $\Pb(T_2^{\ast} \geq \tau_2)$ given by \eqref{Eq:DistT2}, and $\int_{\tau_2}^{\infty}\Pb(T_2^{\ast} > \tau)\di\tau$ given by \eqref{Eq:DblExcur_Part3_i} for $\tau_1 \leq \tau_1^{\ast}$ and by \eqref{Eq:DblExcur_Part3_ii} for $\tau_1 > \tau_1^{\ast}$.

    By substituting \eqref{Eq:DblExcur_Part1}, \eqref{Eq:DblExcur_Part2}, and \eqref{Eq:DblExcur_Part3} into \eqref{Eq:DblExcur_0}, we obtain the probability of the question. For illustration, we present the final result for different possibilities of $\tau_1$ and $\tau_2$
    \begin{itemize}
    \item For $\tau_1 \leq \tau_1^{\ast}$ and $\tau_2 = 0$:
        \begin{equation}\label{Eq:DblExcur_Case1}
        P = \E\{U_{\gamma_1}\} e^{-V (\tau_1^{\ast})^2} \sqrt{\frac{\pi}{4V}}.
        \end{equation}
    \item For $\tau_1 \leq \tau_1^{\ast}$, and $\tau_2 > 0$:
        \begin{equation}\label{Eq:DblExcur_Case2}
        P = \E\{U_{\gamma_1}\} e^{-2 V(\tau_1^{\ast})^2} \frac{e^{-V\tau_2^2}}{e^{-V\tau_1^2}} \Big[\tau_2 + \sqrt{\frac{\pi}{4V}}\frac{\erfc(\sqrt{V}\tau_2)}{e^{-V\tau_2^2}}\Big].
        \end{equation}
    \item For $\tau_1 > \tau_1^{\ast}$ and $\tau_2 \geq \tau_2^{\ast}$, we have $\tau_1^2 \leq \tau_2^2 + (\tau_1^{\ast})^2$, so
        \begin{equation}\label{Eq:DblExcur_Case3}
        P = \E\{U_{\gamma_1}\} e^{-V (\tau_1^{\ast})^2} e^{-V\tau_2^2} \Big[\tau_2 + \sqrt{\frac{\pi}{4V}}\frac{\erfc(\sqrt{V}\tau_2)}{e^{-V\tau_2^2}}\Big].
        \end{equation}
    \item For $\tau_1 > \tau_1^{\ast}$ and $\tau_2 < \tau_2^{\ast}$, we have $\tau_1^2 > \tau_2^2 + (\tau_1^{\ast})^2$, so
        \begin{equation}\label{Eq:DblExcur_Case4}
        P = \E\{U_{\gamma_1}\} e^{-V\tau_1^2} \Big[\tau_2^{\ast} + \frac{e^{-V (\tau_1^{\ast})^2}}{e^{-V\tau_1^2}} \sqrt{\frac{\pi}{4V}} \frac{\erfc(\sqrt{V}\tau_2)}{e^{-V(\tau_2^{\ast})^2}}\Big].
        \end{equation}
    \end{itemize}

\section{Conclusions}

    This paper addressed some properties of level crossings for a stationary normal process. We showed that the length of up-excursions above a large negative level $-\gamma$ is asymptotically exponentially distributed as $-\gamma \to -\infty$. Besides, the simple analytical expression provided by the exponential distribution, this result clarifies the difference between up-excursions above a large negative and above a large positive level in that a stationary normal process stays most of the time above a large negative level, while it stays above a large positive level during short intervals. After that, using the asymptotic parabolic trajectory of large excursions, we derived the mean number of crossings as well as the length of successive excursions above two subsequent large levels. Simulations showed that the asymptotic results are also effective for practical values of crossing levels. We showed an example where the developed theory is applied to derive the probability of successive excursions above subsequent levels during a time window.



\ifCLASSOPTIONcaptionsoff
  \newpage
\fi

\bibliographystyle{IEEEtranTCOM}
\bibliography{IEEEabrv,levelcrossing}
%
%

%





\end{document}